\documentclass[preprint,aps,showpacs,nofootinbib,preprintnumbers,amsmath,amssymb]{revtex4-1}
\usepackage{slashed}
\usepackage{epsfig}
\usepackage{url,hyperref}
\usepackage{dcolumn}
\usepackage{bm}
\usepackage[usenames ,dvipsnames]{xcolor}
\usepackage{slashed}
\usepackage{graphicx,color}
\usepackage{axodraw}
 \usepackage{url}
\begin{document}
\baselineskip 20pt

\title{Hidden Photon Compton and Bremsstrahlung in White Dwarf Anomalous Cooling and Luminosity Functions}

\baselineskip 20pt

\author{
Chia-Feng Chang
}

\affiliation{
Department of Physics, National Taiwan University, Taipei 116, Taiwan
}

\date{\today}                                          

\begin{abstract}

We computed the contribution of the Compton and Bremsstrahlung processes with a hidden light $U(1)_D$ neutral boson $\gamma_D$ to the white dwarf G117-B15A anomalous cooling rate,  as well as the white dwarf luminosity functions (WDLF). We demonstrated that for a light mass of hidden photon ($m_{\gamma_D} \ll$ a few keV), compatible results are obtained for the recent Sloan Digital Sky Survey and the SuperCOSMOS Sky Survey observation, but the stringent limits would be imposed on the kinetic mixing $\epsilon$. We performed $\chi^2$-tests to acquire a quantitative assessment on the WDLF data in the context of our model, computed under the assumption of different kinetic mixing $\epsilon$, the age of the oldest computed stars $T_D$, and a constant star formation rate $\psi$. Then taken together, the WDLF analysis of 2$\sigma$ confidence interval $\epsilon = \left( 0.37^{+0.35}_{-0.37}\right) \times 10^{-14}$ is barely consistent with the cooling rate analysis at 2$\sigma$ regime $\epsilon = \left( 0.97^{+0.35}_{-0.37} \right) \times 10^{-14}$. The two approaches used here agree with each other in yielding an anomalous cooling rate of white dwarf in this luminosity range.

\end{abstract}

\maketitle

\section{Introduction}

The addition of an extra $U(1)_D$ gauge symmetry, implying the existence of an exotic massive neutral gauge boson, is one of the much investigated extensions of the Standard Model (SM) \cite{SM}. This exotic gauge boson is called hidden(dark) photon and denoted by $\gamma_D$ in literatures \cite{Holdom,Nath,HeCalculation,He1,SB,Foot,nonFoot,Hooper1,Hooper2,figure,g-2,My,HZZp-LHC,TC3}, and the interaction between $\gamma_D$ with the Standard Model particles arise only through a kinetic mixing $\epsilon$ \cite{Holdom}. The extra $U(1)_D$ gauge field is predicted in some phenomenology theories \cite{Nath,HeCalculation,He1,SB,Foot,nonFoot,Hooper1,Hooper2,figure,g-2,My,HZZp-LHC,TC3}, in which $\gamma_D$ plays the role of a dark matter candidate. In these theories $\gamma_D$ is assumed to be a product in high energy processes. As a result we can directly or indirectly measure the trace of hidden photon in experiments and observations.

In this paper, we describe the hidden photon production as well as the cooling anomaly in white dwarf. The original hint of a cooling anomaly came from the measurement \cite{OriginalG117}, of the rate of period change $\dot{P}$ for the 215s mode of G117-B15A \cite{G117,G1172}, which is significantly larger than the prediction of standard pulsation theory \cite{Isern1992}, see TABLE.\ref{table1}.

According to previous researches \cite{OriginalG117,Isern1992,G1172,G117}, the relative rate of change of the white dwarf pulsation period $\dot{P}/P$ is essentially proportional to the temperature cooling rate $\dot{T}/T$, which is in accordance with the assumption that the white dwarf (ZZ ceti stars) is not yet crystallized \cite{Isern1992}.  Hence if the anomalous cooling rate as implied by the excess rate of change of the pulsation period are indeed induceed by the hidden photons, one would get\cite{Isern1992,151208108}
\begin{align}
\label{LST}
\frac{L_{\gamma_D}}{L_{st}} \simeq \frac{\dot{P}_{obs}}{\dot{P}_{th}}-1,
\end{align}
where $L_{st}$ is the standard luminosity of the pulsation white dwarf. This relation is available only when $L_{\gamma_D} \geq L_{st}$ from the results reported in the reference research \cite{OriginalG117}. After the original measurement of \cite{OriginalG117}, astrophysicists also observed other two pulsation white dwarfs in two decades. The most present results \cite{151208108} are shown in TABLE.\ref{table1}, in which the data of two DA white dwarfs G117-B15A \cite{G1172,G117}, R548 \cite{R548} and a DB white dwarf PG 1351+489 are presented \cite{PG1351}. As this table, the constraint from G117-B15A is stronger then others.
\begin{table}[b!] 
\setlength{\arrayrulewidth}{0.95pt}
\tabcolsep=4pt
\renewcommand{\arraystretch}{0.2}
\caption{\baselineskip 14pt
Results of $\dot{P}$ for G117-B15A \cite{G117}, R548 \cite{R548}, PG 1351+489 \cite{PG1351}. This table is referenced from \cite{151208108}.} 
\centering 
\begin{tabular}{c c c c c } 
\hline\hline 
WD & class $\quad$ & $\dot{P}_{obs}[\hbox{s/s}]$ & $\quad$ $\dot{P}_{th}[\hbox{s/s}]$ & $\dot{P}_{obs} / \dot{P}_{th}-1$  \\ [1ex]
\hline
G117 - B15A & $\quad$ DA $\quad$ & $(4.19 \pm 0.73) \; \times \; 10^{-15}$ & $\quad$ $(1.25 \pm 0.09) \; \times \; 10^{-15}$ & $2.35 \pm 0.63$ \\ [1ex]
R548  & $\quad$ DA $\quad$ & $(3.33 \pm 1.1) \; \times \; 10^{-15}$ & $\quad$ $(1.1 \pm 0.09) \; \times \; 10^{-15}$ & $2.03 \pm 1.03$ \\ [1ex]
PG 1351+489 & $\quad$ DB $\quad$ & $(2.0 \pm 0.9) \; \times \; 10^{-13}$ & $\quad$ $(0.81 \pm 0.5) \; \times \; 10^{-13}$ & $1.47 \pm 1.89$ \\ [1ex]
\hline\hline 
\end{tabular} 
\label{table1} 
\end{table} 

The first luminosity function was derived almost fifty years ago \cite{firstLF}, and during the long development period, it has been improved with significant amoumt of research works \cite{SSD1,SSD2,SSD3,SuperCOSMOS,Miller,Isern1,Iben,WDLFx,151208108,Wood1992,Salpeter1955,cutLF,OLF}. The recent availability of data are contributed by the Sloan Digital Sky Survey (SDSS) \cite{SSD1,SSD2,SSD3} and the SuperCOSMOS Sky Survey (SSS) \cite{SuperCOSMOS}, and those has noticeably improved the accuracy of the new luminosity functions. A recent analysis of the white dwarf luminosity functions (WDLF) is done by \cite{Miller} in which a unified WDLF is constructed by averaging the SDSS and SSS, and estimated the uncertainties by taking into account the differences between the WDLF at each magnitude bin.  

We use the \texttt{LPCODE} stellar evolution code\footnote{ \texttt{LPCODE} website: \url{fcaglp.fcaglp.unlp.edu.ar/evolgroup}} \cite{LPCODE29} for our stellar evolution computations. This code has been employed to study different problems related to the formation and evolution of white dwarfs \cite{LPCODE28,LPCODE29,LPCODE30,LPCODE0,LPCODEre}. In the following paragragh, we briefly outline the algorithm work of \texttt{LPCODE}, and more details are given by \cite{LPCODE29,14067712}. Radiative opacities are those of \cite{LPCODE31} while conductive opacities are cited by \cite{LPCODE32}, complemented at low temperatures (molecular opacities), which are produced by \cite{LPCODE33}. The equation of state for the high density regime is cited by \cite{LPCODE34}, while for the low density regime, an updated version of the equation of state of \cite{LPCODE35} is used. Neutrino cooling by pair, photo-, plasma, Bremsstrahlung production are included following the results of \cite{LPCODE36}, while plasma processes are also included \cite{LPCODE37}. White dwarf models computed with \texttt{LPCODE} also include detailed non-gray model atmospheres \cite{LPCODE38}. In addition, the effects of time dependent element diffusion during the white dwarf evolution are following by \cite{LPCODE39} for multicomponent gases. Finally, we have to mention that \texttt{LPCODE} has been tested against with other white dwarf evolutionary codes, and the discrepancies are lower than 2\%, see \cite{LPCODE40}.

In Sec.$\;$2, we set up our model and also briefly discuss the mixing between the three neutral gauge bosons in the model as studied previously \cite{Nath,My}. In Sec.$\;$3, we present the calculations of hidden photon Compton scattering and Bremsstrahlung processes on white dwarf conditions. In Sec.$\;$4, we study that the effect contributed to hidden photon on white dwarf anomalous cooling rate and WDLF. We conclude our results in Sec.$\;$5.

\section{Hidden Photon Model}
In this section we discuss the $U(1)_D$ extension of the Standard Model(SM) with a gauge kinetic mixing. We assume that the all of SM particles do not carry $U(1)_D$ quantum numbers, and non-SM particles do not carry quantum numbers of the SM gauge group. The Lagrangian describing the coupled system is 
\begin{align}
\label{Lgauge}
\mathcal{L}_B = -\frac{1}{4} B_{\mu \nu} B^{\mu \nu} -\frac{1}{4} C_{\mu \nu} C^{\mu \nu} - \frac{m^2}{2}C_{\mu}C^{\mu}  - \frac{\epsilon}{2} B_{\mu \nu} C^{\mu \nu}.
\end{align}
In additional to the mass mixing of the three neutral gauge boson arise from the spontaneously electroweak symmetry breaking without term of kinetic mixing given by
\begin{align}
\mathcal{L}_m = \frac{1}{2} ( C^{\mu} \;\; B^{\mu} \;\; W^{3 \mu} ) \; M^2 \; 
\left(
\begin{tabular}{c}
$C_{\mu}$
\\
$B_{\mu}$
\\
$W^3_{\mu}$
\end{tabular}
\right)
\end{align}
with the following mass mixing matrix
\begin{equation}
M^2 = \left(
\begin{tabular}{ccc}
$m^2$ \; & \; $0$ 
\; & \; $0$ \\
$0$ \; & \; $\frac{1}{4}g_Y^2 v^2$ 
\; & \; $ -\frac{1}{4}g g_Y v^2$ \\
$0$ \; & \; $-\frac{1}{4}g g_Y v^2$ 
\; & \; $\frac{1}{4}g^2 v^2$
\end{tabular}
\right) \; .
\end{equation}
We also have the kinetic mixing between the two U(1)s from the last term in Eq.~(\ref{Lgauge}). Both
the kinetic and mass mixing can be diagonalized simultaneously by the following mixed
transformation 
\begin{equation}
\left(
\begin{tabular}{c}
$C_\mu$ 
\vspace{0.05in}
\\
$B_\mu$ 
\vspace{0.05in}
\\
$W^3_\mu$
\end{tabular}
\right) =
K \cdot O
\left(
\begin{tabular}{c}
$\gamma_{D,\mu}$
\vspace{0.05in}
 \\
$Z_\mu$
\vspace{0.05in}
 \\
$A_\mu$
\end{tabular}
\right) \; ,\;\;\;\;\;\;\;
K = \left(
\begin{tabular}{ccc}
$\beta$ & 0 & 0 \\
$-\epsilon \beta$ & 1 & 0 \\
0 & 0 & 1
\end{tabular}
\right) \; , 
\end{equation}
where $\gamma_{D}$ , $Z_μ$ and $A_μ$ are the physical hidden photon, $Z$ boson and the photon respectively. The matrix $K$ is a general linear transformation that diagonalizes the kinetic mixing, and the parameter $\beta =1/\sqrt{1-\epsilon^2}$ with $\epsilon < 1$, and $O$ is a 3 $\times$ 3 orthogonal matrix which can be parametrized as
\begin{equation}
 \left(
\begin{tabular}{ccc}
$\cos\psi \cos \phi - \sin \theta_w \sin \phi \sin \psi$ \; & \; $\sin \psi \cos \phi + \sin \theta_w \sin \phi \cos \psi$ 
\; & \; $- \cos \theta_w \sin \phi$ \\
$\cos\psi \sin \phi + \sin \theta_w \cos \phi \sin \psi$ \; & \; $\sin \psi \sin \phi - \sin \theta_w \cos \phi \cos \psi$ 
\; & \; $ \cos \theta_w \cos \phi$ \\
$- \cos \theta_w \sin \psi$ \; & \; $\cos \theta_w \cos \psi$ 
\; & \; $\sin \theta_w $
\end{tabular}
\right)
\end{equation}
with the mixing angles defined as
\begin{equation}
\label{anglet}
\tan\theta_w = \frac{g_Y}{g}\, , \quad \tan\phi = \frac{-\epsilon}{\sqrt{1-\epsilon^2}}\, ,
\quad   \tan\psi = \pm \frac{\tan\phi \cos\theta_w}{\tan\theta_w} \left[\frac{1-m^2_Z/M^2_W}{1-m^2_Z/m^2} 
+ \tan^2\theta_w \right]  \; .
\end{equation}
After the $K$ transformation, the gauge bosons mass matrix is
\begin{equation}
{\tilde M}^2 = K^T \cdot M^2 \cdot K = 
\left(
\begin{tabular}{ccc}
$\beta^2( m^2 + \frac{1}{4} \epsilon^2 g_Y^2 v^2)$ & $-\frac{1}{4}\epsilon \beta g_Y^2 v^2$ & $\frac{1}{4} \epsilon \beta g g_Y v^2$  \\
$- \frac{1}{4} \epsilon \beta  g_Y^2 v^2$ \; & \;
$\frac{1}{4} g_Y^2 v^2$ \; &
$- \frac{1}{4}g g_Y v^2$ \\
 $- \frac{1}{4} \epsilon \beta g g_Y v^2$ & $\frac{1}{4}g g_Y v^2$ & $ \frac{1}{4} g^2 v^2 $
\end{tabular}
\right) \; .
\end{equation}
The $O$ matrix diagonalize this ${\tilde M}^2$ matrix 
\begin{equation}
M^2_{Diag} = O^T \cdot {\tilde M}^2 \cdot O = 
\left(
\begin{tabular}{ccc}
$m^2_{\gamma_D}$ & 0 & 0 \\
0 & $m^2_Z$ & 0 \\
0 & 0 & $m^2_\gamma$
\end{tabular}
\right) \; ,
\end{equation}
with the following eigenvalues (assuming $m_{\gamma_D} \leq m_Z$)
\begin{equation}
m^2_{\gamma}=0, \; \; \; \; m^2_{Z,\gamma_D} = (q \pm p)/2 
\end{equation}
where
\begin{equation}
p = \sqrt{q^2- m^2 v^2 (g^2+g_Y^2) \beta^2} \; ,\;\;\;\;\; q = m^2 \beta^2 + \frac{1}{4} (g^2 +g_Y^2 \beta^2)v^2 \; .
\end{equation}
For small kinetic parameter mixing $\epsilon$, the $Z$ and $\gamma_D$ masses can be approximated by $m_Z \approx \sqrt{(g^2+g_Y^2)v/2}$ and $m_{\gamma_D} \approx m$. It is noteworthy that the mass matrix is calculated in \cite{HeCalculation,Nath,My}.
The couplings of these physical neutral gauge bosons with the SM fermions, we refer the readers to Eq.~(11) in \cite{Nath}.
\begin{align}
\label{Lint}
\mathcal{L}_{int} = \frac{\sqrt{g^2 + g_Y^2}}{2}  \bar{f} \gamma^\mu [ (v^\prime_f - \gamma_5 a^\prime_f )\gamma_{D,\mu} + (v_f - \gamma_5 a_f ) Z_\mu ] f + e \bar{f} \gamma^\mu Q_f A_\mu f,
\end{align}
with Eq.~(27-30) in \cite{Nath}, and choose the minus solution at Eq.~(\ref{anglet}). As above, we have the relation when $m_Z \gg m_{\gamma_D} $ and $1 \gg \epsilon $
\begin{align}
\tan\psi \approx \sin\psi \approx - \epsilon \sin\theta_w ,
\end{align}
and also
\begin{align}
v^\prime_f & \approx 2 \sin\theta_w \epsilon (1-\sin^2 \theta_w) Q_f,\;\;\;\;\;\;
a^\prime_f \approx 0.
\end{align}
Further, the interaction term of hidden photon is
\begin{align}
\mathcal{L}_{int} = e \cos \theta_w \bar{f}\gamma^\mu f \gamma_{D,\mu}. 
\end{align}
where the Weinberg angle $\cos \theta_w = 0.872$. 

\section{Calculations}
For all calculations we start from the energy loss(cooling) rate \cite{Raffelt}
\begin{align} \label{Q}
\notag Q = \prod^{N_i}_{i=0} \int {d^3 p_i \over 2 E_i (2 \pi)^3 } f_i(E_i) \prod^{N_f}_{f=0}  & \int {d^3 p_f \over 2 E_f (2 \pi)^3 } [1 \pm f_f(E_f)]  \int {d^3 p_{\gamma_D} \over 2 E_{\gamma_D} (2 \pi)^3 } E_{\gamma_D} (1 + f_{\gamma_D}) \\
\times & {1 \over N_{i,id}!} {1 \over N_{f,id}!} \sum_{s, p} |\mathcal{M}|^2  (2 \pi)^4 \delta^4(\sum^N_{i=1}p_i-\sum^N_{f=1} p_f - p_{\gamma_D}).
\end{align}
The subscript of $i$ and $f$ are meaning initial state and final state, respectively, and the $f$ is the distribution function.  The $f_{\gamma_D}$ is limited to zero, because a small density of dark photon in space is assumed. The minus sign at front of Fermi distribution function $f_f$ is because of Pauli Blocking effect. In this paper, the anomalous cooling rates are due to the dark photon Compton and Bremsstrahlung processes.

\subsection{Compton Scattering} 
First of all, we consider Compton scattering,
\begin{align}
 \gamma +e \to e + \gamma_D. 
\end{align}
The Compton-type processes are typically important when the electrons are non-degenerate (otherwise bremsstrahlung electrons dominates) and non-relativistic (otherwise $e^+ e^-$ annihilation dominates) \cite{Raffelt}. In this subsection we provide a result of hidden photon Compton processes in stars. In our calculations, we consider the case of white dwarf core
\begin{align}
\label{case22}
m_e \gg T \approx \omega > m_{\gamma_D}, 
\end{align}
where $T$ is the temperature in the core of white dwarf and $\omega$ is a energy of initial photon. The cross section can be written as
\begin{align}\label{ComptonCrossSection}
\sigma \approx 
& \frac{16\pi \alpha^2 \epsilon^2 \cos^2\theta_w}{m_e^2 \omega} \sqrt{\omega^2-m_{\gamma_D}^2}, 
\end{align}
by according to the Eq.~(\ref{Q}), we have the energy loss rate,
\begin{align} \label{ComptonQ}
\notag Q_{Compton} =\; \frac{n_e T^4}{\pi^2}F(k_F) \int^{\infty}_{0} dx \frac{\sigma x^3}{e^x-1},
\end{align}
where $x=\omega/T$, the $k_F$ is Fermi momentum and $F(k_F)$ is given by \cite{Raffelt,14067712}
\begin{align}
F(k_F)= \left\{
\begin{aligned}
& 1, \;\;\;\;\;\;\;\;\;\;\;\;\;\;\;\;\;\;\;\;\;\;\;\;\;\;\;\;\;\;\;\; \;\;\;\;\; \;\;\;\;(\hbox{nondegenerate, non-relativistic}),  \\
& 3E_F T /k_F^2, \;\;\;\;\; \;\;\;\;\; \;\;\;\;\;\;\;\;\; \;\;\;\;\; \;\;\;\;  (\hbox{degenerate}),\\
&  (1+(3E_F T /k_F^2)^{-2})^{-1/2}, \;\;\;\;\;\;\;   (\hbox{combination}),  \\
\end{aligned}
\right.
\end{align}
where $n_e = k_F^3/3\pi^2$.
Following the Eq.~(\ref{ComptonCrossSection}), Eq.~(\ref{ComptonQ}) and the case of white dwarf Eq.~(\ref{case22}), we have the Compton $\gamma_D$ cooling rate,
\begin{align}
Q_{Compton}
 \approx n_e \frac{3 \times 2^5 \zeta_4 \alpha^2 \epsilon^2 \cos^2\theta_w T^3}{\pi m_e^2}\sqrt{T^2-m_{\gamma_D}^2} F(k_F), 
\end{align}
and the cooling rate per unit mass is
\begin{align}
\label{Comptone}
\epsilon_{Compton} = \frac{Q_{Compton}}{\rho} \approx & 5.71 \times 10^{28} \; Z \epsilon^2 \cos^2\theta_w T_8^3 \sqrt{T_8^2-m_{\gamma_{D_8}}^2} F(k_F) \;\; (\hbox{erg} \; \hbox{g}^{-1} \; \hbox{s}^{-1}),
\end{align}
where $\zeta_n$ is Riemann zeta function, $Z$ is the atomic number, and $T_8 = T \times 10^{-8} $ K.
\subsection{Bremsstrahlung}

In this subsection, we consider the collision between an electron and an infinitely heavy nucleus $Ze$,
\begin{align}
e^- (k_1) + Ze(\textbf{k}) & \to  e^-(k_2) + Ze(\textbf{k}) + \gamma_D(k^\prime), 
\end{align}
because the white dwarf electron sea is highly degenerate; therefore the relation between $m_e$ and Fermi momentum($k_F$) of electron gas is $m_e \approx k_F \gg$ Temperature($T \approx$ keV). It is worth to mention that the two electrons bremsstrahlung is suppressed by the Pauli Blocking effect, so that we only have to consider the collision of a electron with an infinitely heavy nucleus. In a highly degenerate case,
\begin{align}
\label{DeCaseLimit}
m_e \approx k_F \gtrsim k_1 \approx k_2 \gg E_{\gamma_D} ,
\end{align}
the small mass of hidden photon ($m_{\gamma_D} \ll$ keV) is also assumed.
For the phase space, we adopt the calculation methods as given in chapter.II of \cite{Flowers1973}, and also replace the plasma propagator as \cite{Itoh1,Itoh2},
\begin{align}
\frac{1}{\textbf{q}^4} \to \frac{S(\textbf{q})|f(\textbf{q})|^2}{|\textbf{q}^2 \epsilon(\textbf{q})|^2},
\end{align}
where $S(\textbf{q})$ is the static structure factor, $f(\textbf{q})$ is an atomic form factor and $\epsilon(\textbf{q})$ is the longitudinal dielectric function of the electron liquid. As above, we have
\begin{align}
\label{QBre2}
Q_{Bre} =  \sum_j n_j \frac{Z_j^2 e^6 \epsilon^2}{2^5 \pi^5} \int^{1}_{0} dq 
\frac{S(\textbf{q})|f(\textbf{q})|^2}{|\textbf{q} \epsilon(\textbf{q})|^2}
(I_a(\textbf{q})+I_b(\textbf{q})) 
\int \frac{E_{\gamma_D} dE_{\gamma_D}}{e^{E_{\gamma_D}/T}-1},
\end{align}
with
\begin{align}
\notag I_a(\textbf{q}) =  ( r_{m_f}^2 + 1 -r_q^2/4 ) \int^1_{-1} I_1(\cos\theta,1-r_q^2/2) d\cos\theta,\\
\notag I_b(\textbf{q}) =  ( r_{m_f}^2 + 1 -r_q^2/4 ) \int^1_{-1} I_2(\cos\theta,1-r_q^2/2) d\cos\theta,
\end{align}
where 
\begin{align}
\notag I_1(C,C_e) & = -2\frac{1-r_F^2}{(1- r_F C)(1- r_F (C C_e + S S_e))},\\ 
\notag I_2(C,C_e) & = r_m^2 (\frac{1}{(1-r_F C)^2}+\frac{1}{(1- r_F (C C_e + S S_e))^2}),
\end{align}
the parameters $S = \sqrt{1-C^2}$, $r_F = k_F/E_F$, $r_{m} = m_e/E_F$, $r_{m_f} = m_e/k_F$, $r_q = q/k_F$ and $j$ is the sum over all of possible nuclear in the star. In the plasma part, we assume the core of white dwarf is almost a one-component plasma. The structure factor can be found in Eq.~(53) of \cite{condmat9810213,98110529903127}, also the small \textbf{q} limit is
\begin{align}
S(\textbf{q}) \approx \frac{1}{(\kappa_D/\textbf{q})^2 + \beta_q + \mathcal{O}(\textbf{q}^2)},
\end{align} 
and the function $\beta_q$ is shown in Eq.~(56) and Eq.~(59) of \cite{condmat9810213}.
The form factor $f(\textbf{q})$ and dielectric function $\epsilon(\textbf{q})$ can be found in \cite{LPCODE37,Itoh1,Itoh2,J1962NV1973}.
For the numerical results, we find the integrations can be approximated to a function form, but we have to note that the approximations are available only in range of $10^{-2} \leq T_8 \leq 10^2$. The size of errors in this approximation is less than about 15\%(1\%) in case of $T_8 \gtrsim 4(\lesssim 4)$,
\begin{align}
\notag \int^{1}_{0} dq 
\frac{S(\textbf{q})|f(\textbf{q})|^2}{|\textbf{q} \epsilon(\textbf{q})|^2}
(I_a(\textbf{q})+I_b(\textbf{q})) \approx \left\{
\begin{aligned}
& C^{i}_1 \times T_8^{0.31}, \;\;\;\;\;\;\;\;\;\;\;\;\;\;\;\;\;\;\;\;( \rho_6 \approx 1; T_8 \gtrsim 4),  \\
& C^{i}_2 \times T_8^{0.85}, \;\;\;\;\;\;\;\;\;\;\;\;\;\;\;\;\;\;\;\;( \rho_6 \approx 1; T_8 \lesssim 4) ,  \\
\end{aligned}
\right.
\end{align}
with
\begin{align}
C^{\hbox{O}}_1 = 9.0, \;\; C^{\hbox{O}}_2 = 4.2, \;\; C^{\hbox{C}}_1 = 11, \;\; C^{\hbox{C}}_2 = 6.2,
\end{align}
where $i= \hbox{O},\hbox{C}$. Those are corresponded to oxygen and carbon, respectively. Finally, the Eq.~(\ref{QBre2}) can be written as
\begin{align}
\label{QBre3}
\notag \epsilon_{Bre} =
 7.06 \times 10^{27} \sum_j  \frac{Z_j^2 \epsilon^2 \cos^2\theta_w T_8^2}{  A_j} \int^{1}_{0} dq 
\frac{S(\textbf{q})|f(\textbf{q})|^2}{|\textbf{q} \epsilon(\textbf{q})|^2}
(I_a(\textbf{q})+I_b(\textbf{q})) (\hbox{erg} \; \hbox{g}^{-1} \; \hbox{s}^{-1} ) . 
\end{align}
In white dwarf core, the abundance is large dominate by oxygen and carbon, therefore we can only obtain the numerical results of cases of oxygen and carbon,
\begin{align}
\epsilon^{i}_{Bre} \approx & \left\{
\begin{aligned}
& R^i_1 \times 10^{29} \;\;  \epsilon^2 \cos^2\theta_w T_8^{2.31} \;\; (\hbox{erg} \; \hbox{g}^{-1} \; \hbox{s}^{-1} ), \;\;\;\;\;\;\;\;\;\;\;\;\;\;\;\;(\rho_6 \approx 1; T_8 \gtrsim 4),  \\
& R^i_2 \times 10^{29} \;\;  \epsilon^2 \cos^2\theta_w T_8^{2.85} \;\; (\hbox{erg} \; \hbox{g}^{-1} \; \hbox{s}^{-1} ), \;\;\;\;\;\;\;\;\;\;\;\;\;\;\;\;(\rho_6 \approx 1; T_8 \lesssim 4) ,  \\
\end{aligned}
\right.
\end{align}
with
\begin{align}
R_1^{\hbox{O}} = 2.54, \;\; R^{\hbox{O}}_2 = 1.17,
 \;\; R^{\hbox{C}}_1 = 2.34, \;\; R^{\hbox{C}}_2=1.31.
\end{align}

\section{Phenomenology}

\subsection{White Dwarf Cooling Rate}

In this section, we focus on the cooling rates of the pulsating white dwarf star G117-B15A\cite{0711,151208108,G117,R548, WD13,Isern1,0104103,Isern1992}.
The rates of period change for G117-B15A do not take into account other energy source than gravothermal energy for the evolutionary cooling of the star. In order to explain the rates of period change (see, Table.\ref{table1}), the hidden photon would carry some of energy out from the white dwarf, therefore it affects the rates of period change. For the numerical analysis, the Eq.~(\ref{LST}) is used, the value of $L_{st}$ is taken by $1.2 \times 10^{31}$ erg/s \cite{G117,151208108,LPCODE29}, and the observation results are shown in Table.\ref{table1}. We also take the abundance $X_C$ and $X_O$ as\cite{0711,G117,LPCODE29}
\begin{align}
X_C \approx 0.3, \;\;\;\; X_O \approx 0.7.
\end{align}
\begin{figure}[t!]
\centering
\includegraphics[width=0.45\textwidth]{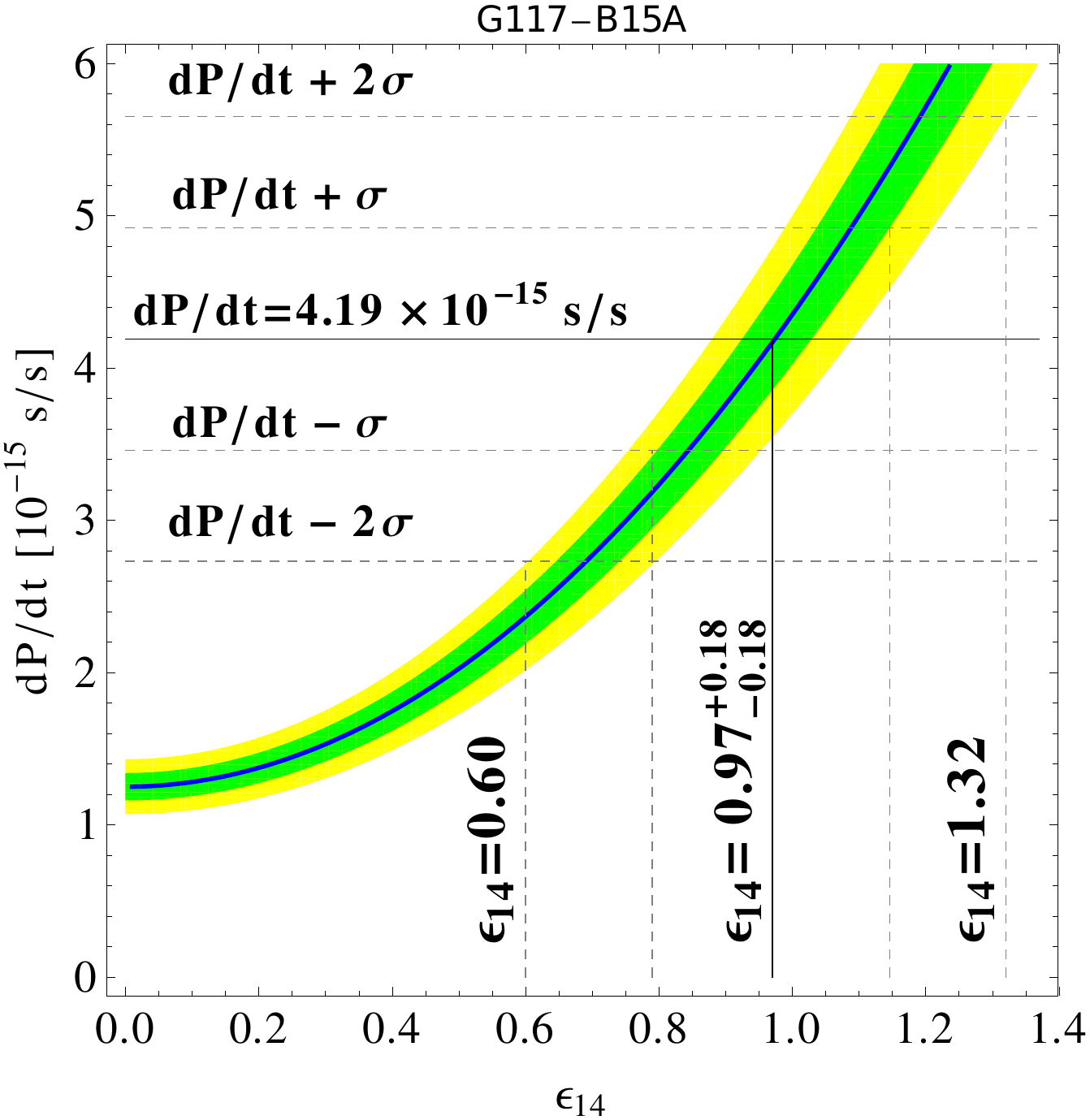} 
\caption{\small \label{G117B15AFig}
\baselineskip 17pt
The rate of period change $\dot{P}$ in terms of hidden photon kinetic mixing parameter $\epsilon$, the green line indicate the Eq.~(\ref{Pst}), and the green(yellow) regime is the error of formula with 1$\sigma$(2$\sigma$). The horizontal black line indicates the observed value with its corresponding uncertainties \cite{G1172,G117,0711}. As the Figure, the amount of kinetic mixing $\epsilon= (0.97^{+0.18}_{-0.18}) \times 10^{-14}$ is at 1$\sigma$ level, and the amount of kinetic mixing $\epsilon= (0.97^{+0.35}_{-0.37}) \times 10^{-14}$ is at 2$\sigma$ level.}
\end{figure}
For a mixture of carbon and oxygen, the hidden photon emission rate is given by previous section.
The information of G117-B15A can be found in \cite{G117} and \texttt{LPCODE} \cite{LPCODE29}.
As above, we have the cooling rate of G117-B15A as
\begin{align}
\label{Pst}
\dot{P}_{st+\gamma_D} \times 10^{15} \approx (2.11 \pm 0.16) \epsilon_{14}^2 \cos^2\theta_w T_7^{2.85} +1.25 \pm 0.09  \;\; \hbox{s/s}
\end{align}
where $\epsilon_{14} = \epsilon \, \times \, 10^{14}$. In Fig.~\ref{G117B15AFig}, we display the theoretical value of $\dot{P}$ as Eq.~(\ref{Pst}), and the observed value $\dot{P}=4.19 \times 10^{15}$ s/s for increasing values of the kinetic mixing $\epsilon$ (blue solid curve), and the uncertainty is estimated in green(1$\sigma$) and yellow(2$\sigma$) area. In this figure, the amount of kinetic mixing $\epsilon_{14}$ is $0.97^{+0.18}_{-0.18}$ in one standard deviation from the observational value. We assume that the anomalous rate of cooling of white dwarf is entirely due to hidden photon, therefore the result is an indirect measurement. But if there existed some of other mechanism in white dwarf, the result will appear merely an upper bound. According to this result, the value of coupling is extremely small, so the hidden photon cannot be observed in collision experiments.  

\subsection{White Dwarf Luminosity Function}

The numerical method of theoretical WDLF is introduced by \cite{Iben}. In this section, we rely on their method, and consider a simplest case of conditions of WDLF. A detailed explanation of the method can be found in \cite{Iben}. According to previous research works \cite{Iben, Wood1992, Isern1,Salpeter1955}, the number of white dwarfs per logarithmic luminosity and volume is 
\begin{align}
\label{LF1}
\frac{dn}{dl} = - \int^{M_2}_{M_1} \psi(t) \left(\frac{dn}{dM_i}\right) \left(\frac{\partial t_c}{\partial l}\right)_m dM_i,
\end{align}
where $\psi(t)$ is the Galactic stellar formation rate at time $t$, $n(M_i)$ is the initial mass function and $t_c(l,M_f)$ is the time since the formation of the white dwarf, of mass $M_f$, for the star to reach a luminosity $log(L/L_{sun}) = l$. In Eq.~(\ref{LF1}), the initial-final mass relation $M_f(M_i)$ and the pre-white dwarf stellar lifetime $t_{ev}(M_i)$ are taking by \cite{Iben,Salpeter1955}. The white dwarf luminosity $l$ and mass of the progenitor $M_i$, the formation time of the star $t$ is obtained by solving 
\begin{align}
\label{LFt}
t+t_{ev}(M)+t_c(l,m)=T_D,
\end{align}
where $T_D$ is the age of the oldest computed stars. The lowest initial mass that produces a white dwarf with luminosity $l$ at the present time. The $M_1$ is obtained from Eq.~(\ref{LFt}) when $t=0$. The value of $M_2$ corresponds to the largest stellar mass progenitor that produces a white dwarf. A constant star formation rate(SFR) is adopted as \cite{Iben, Wood1992, Isern1}
\begin{align}
\psi(t) = \hbox{constant}.
\end{align}

\begin{figure}[t!]
\centering
\includegraphics[width=0.65\textwidth]{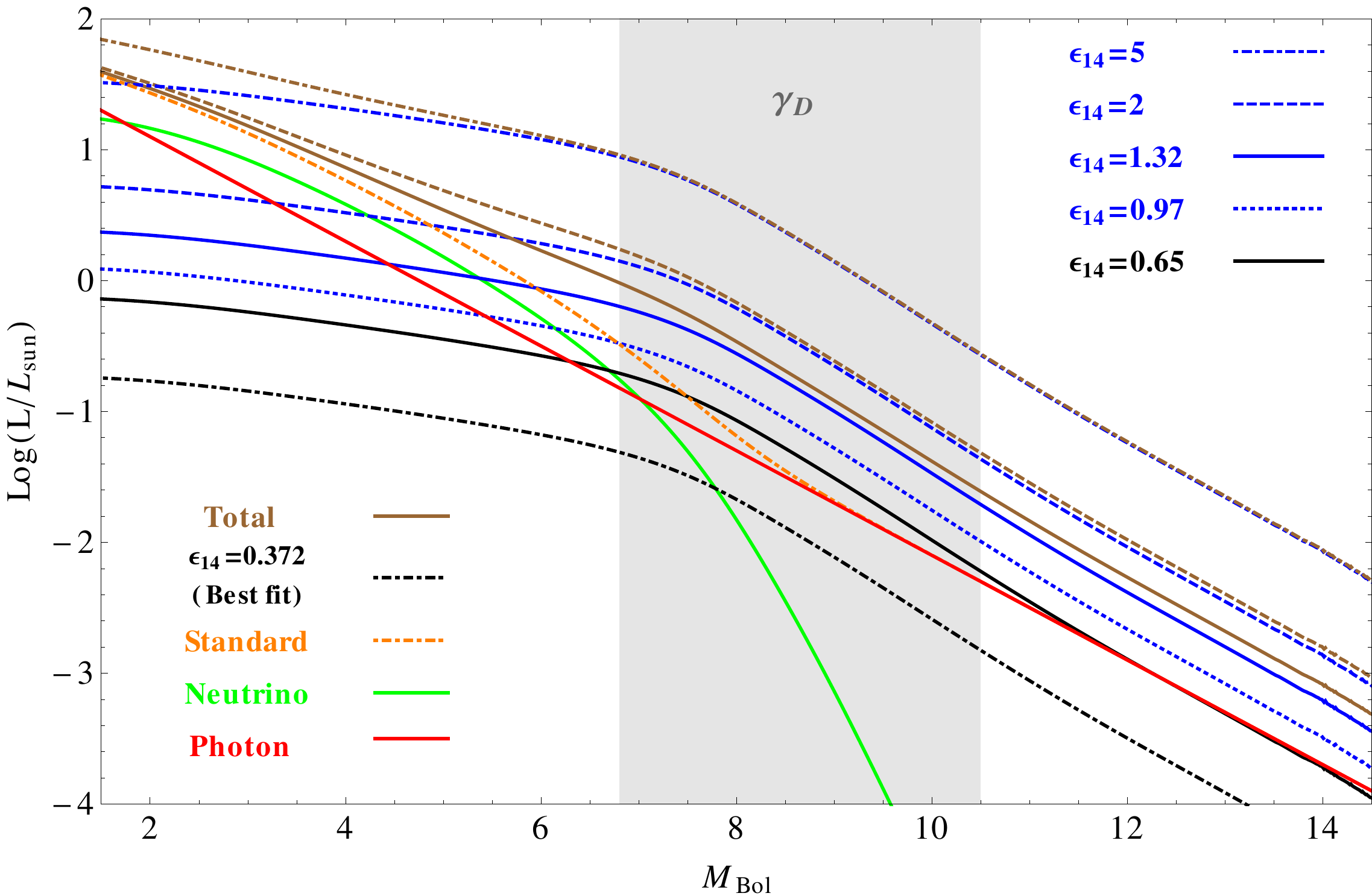} 
\caption{\small \label{LuminosityFunctionFig1}
\baselineskip 16pt
Photon (red curve), neutrino (green curve) and hidden photon (black and blue curves) emission for the mass 0.609$M_{sun}$ at different kinetic mixing parameter $\epsilon$. The impact of hidden photon emission in the core of the white dwarf is dominant in range of $ 6.8 \leq M_{Bol} \leq 10.5$ (shadow) if $\epsilon_{14} \approx 0.65$ (black solid curve), and our best $\chi^2$ fitting of WDLF is $\epsilon_{14}=0.372$ (black dot-dashed curve). The particles emission without hidden photon is shown by orange dot-dashed line. Sum over all of particles emission is taking the form of brown line. 
}
\end{figure}  

Fig.~\ref{LuminosityFunctionFig1} shows the photon, neutrino and hidden photon emissions for the white dwarf with the mass $0.609M_{sun}$ under the assumption of different kinetic mixing parameter $\epsilon$. In particular, the emission of hidden photon cannot be neglected in the range of $ 6.8 \leq M_{Bol} \leq 10.5$ when $\epsilon_{14} \gtrsim 0.65$.  As the comparison of hidden photon emission (black or blue curves) with the standard emission (orange curve, without hidden photon), the result showed that when hidden photon is included, this leads to an extra cooling of the white dwarf core that alters the thermal structure of the white dwarf (shadow). In our best fitting result (black dot-dashed curve), the kinetic mixing parameter $\epsilon_{14}=0.372$ is within a 1.64$\sigma$-like significance level, and the photon is dominant in range of $ M_{Bol} \geq 6.8$. The results indicate that at lower luminosities ($M_{Bol} \gtrsim 6.8$), the neutrino emission becomes negligible and the hidden photon support a small effect on white dwarf luminosity. On the other hand, the best fitting point is same with standard model estimations in the range of  $ M_{Bol} \leq 6.8$.

\begin{figure}[t!]
\centering
\includegraphics[width=0.65\textwidth]{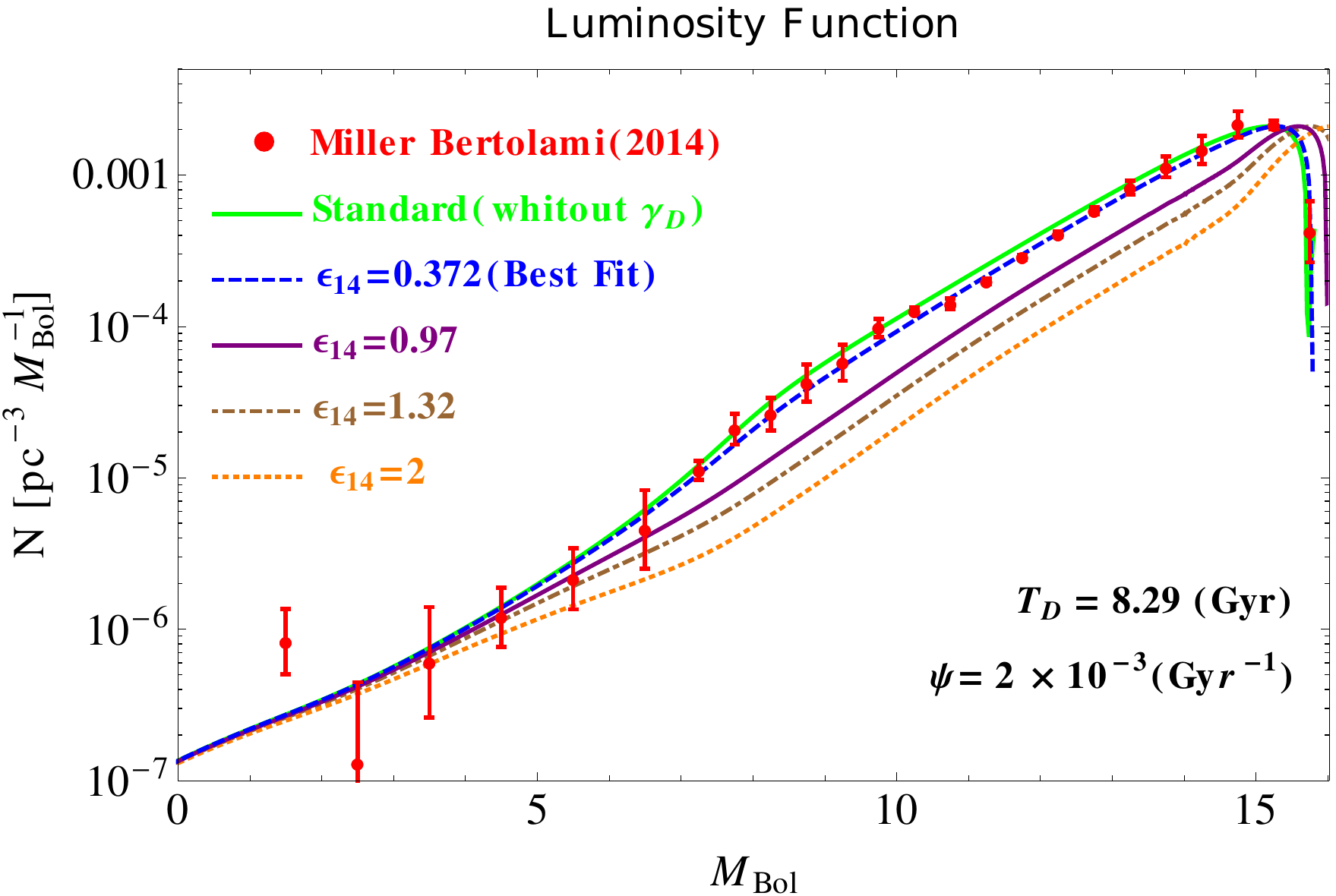} 
\caption{\small \label{LuminosityFunctionFig2}
\baselineskip 16pt
Comparison of the WDLF from \cite{Miller} with the theoretical WDLF on different kinetic mixing parameter $\epsilon$. The Best $\chi^2$ fitting is within a 1.64$\sigma$-like significance level.
}
\end{figure}

Fig.~\ref{LuminosityFunctionFig2} shows a comparison between the theoretical WDLF computed for different kinetic mixing parameter $\epsilon$ and the WDLF of the Galactic disk constructed by \cite{Miller}. This WDLF was constructed from the WDLF determined by the \cite{SSD1,SSD2,SSD3} and the \cite{SuperCOSMOS} sky surveys. The size of the error bars reflects the discrepancies between both WDLFs, see \cite{Miller} for a detailed discussion on these issues. According to the previous work \cite{OLF}, the luminosity function clearly proves that the evolution of white dwarfs is a simple gravothermal process, while the sharp cut-off at high $M_{bol}$ is the consequence of the finite age of the galaxy \cite{cutLF,Isern1}. 

Fig.~\ref{LuminosityFunctionFig2} clearly demonstrated the process of the fitting. Here we fix the age of the oldest computed stars $T_D = 8.29$(Gyr) and a constant star formation rate $\psi = 2 \times 10^{-3}$Gy$\hbox{r}^{-1}$ with $\epsilon_{14}=0.372$ (blue dashed curve). We can find that the standard emission (green) and hidden photon emission (blue-dashed) are separated in range of $M_{Bol} \geq 6.8$, it is meaning hidden photon hold a small portion of the white dwarf luminosity, that effect is not strong. This is corresponded the hidden photon emission ($e.g.$ the black dot-dashed curve) in Fig.~\ref{LuminosityFunctionFig1}.

\begin{figure}[t!]
\centering
\includegraphics[width=0.65\textwidth]{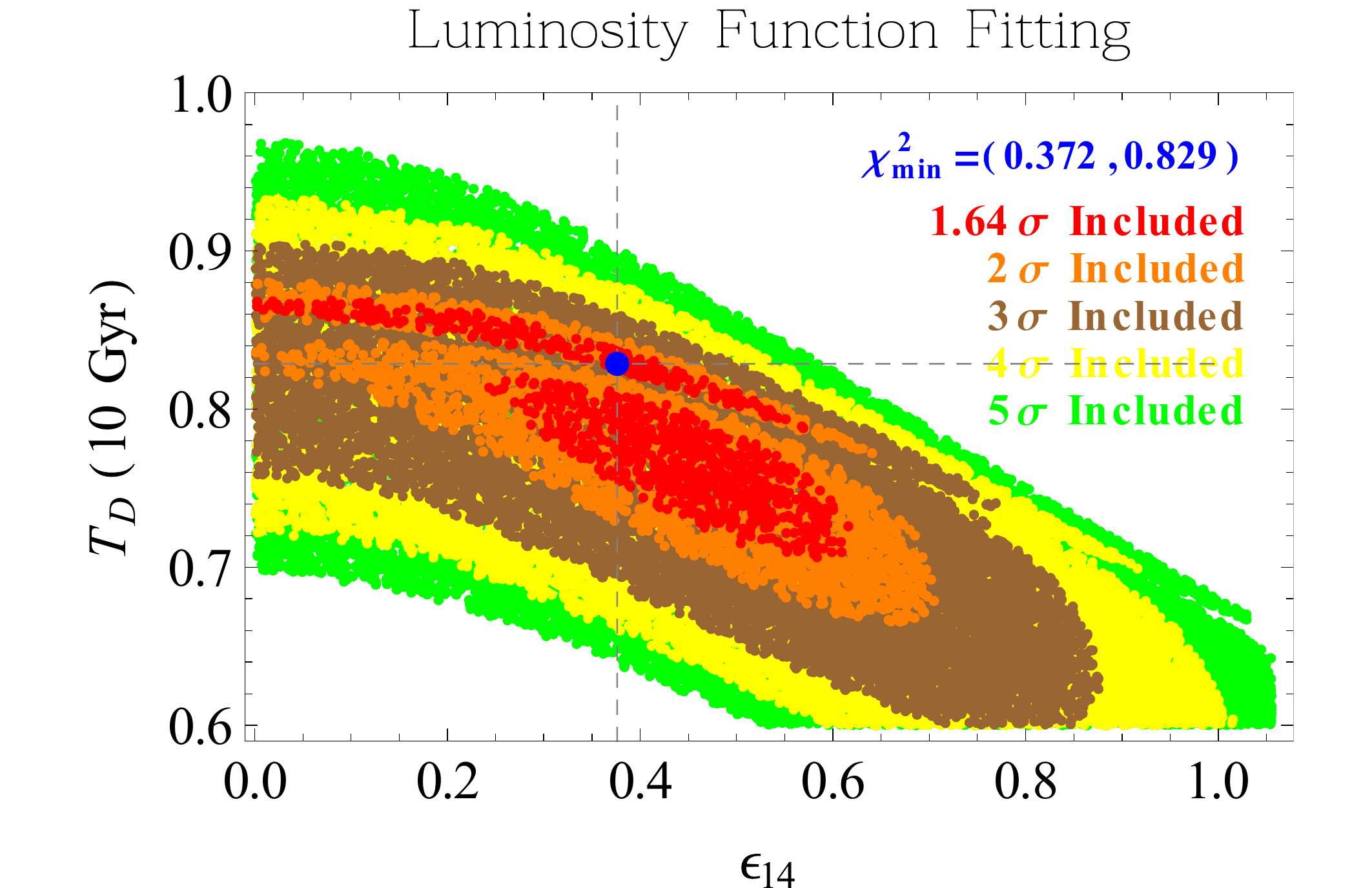} 
\caption{\small \label{LuminosityFunctionFig3}
\baselineskip 16pt
The $\chi^2$ fitting is used in our numerical analysis. We have three free parameters kinetic mixing $\epsilon$, the age of the oldest computed stars $T_D$ and a constant star formation rate $\psi$, and there have 24 data points, therefore our degree of freedom($d.o.f.$) is 20. The best fitting point is $\chi^2_{min}/d.o.f=1.216$ with $\epsilon_{14}=0.372$, $T_D=8.29$Gyr and $\psi=2\times 10^{-3}$(Gy$\hbox{r}^{-1}$). As the result, while all modes below $\epsilon_{14} \leq 0.72$ can't be rejected at a 2$\sigma$-like significance level.
}
\end{figure}

Fig.~\ref{LuminosityFunctionFig3} shows the result of $\chi^2$ fitting, it also express the area of a 1.64$\sigma$ to 5$\sigma$ like significance levels. This figure is separated to upper area and lower area ($e.g.$ red part), because we treat the $\psi$ (SFR) as a free parameter in our fitting, so that the upper side is shown by $\psi \approx 2 \times 10^{-3}$Gy$\hbox{r}^{-1}$ and the lower side is shown by $\psi \approx 2.4 \times 10^{-3}$Gy$\hbox{r}^{-1}$. As the result of orange area, the WDLF constructed with $\epsilon_{14} \leq  0.72$ can not be rejected at more that a 95\% confidence level ($i.e. \approx 2 \sigma$-like), which is barely within the two standard deviations of white dwarf G117-B15A cooling rate, see Fig.~\ref{G117B15AFig}.

\section{Conclusion}

In this paper we have derived an improved value of the kinetic mixing parameter $\epsilon$, assuming that the mass of the hidden photon $\gamma_D$ is smaller than the core temperature of white dwarfs ($\approx$ keV), and also assuming that the enhanced rate of cooling of the pulsating white dwarf is entirely due to the emission of hidden photons. In our calculations, we adopt the stellar evolution code \texttt{LPCODE} \cite{LPCODE29}, and ignore the density of hidden photons in white dwarf core, see Eq.~(\ref{Q}). Based on our assumptions, we found that the anomalous cooling rate of G117-B15A indicates the existence of an additional cooling mechanism in the pulsating white dwarf, consistent with kinetic mixing parameter $\epsilon = (0.97^{+0.18}_{-0.18}) \times 10^{-14}$, see Fig.~\ref{G117B15AFig}. In addition, we analyzed the contribution of hidden photons to the WDLF, see Fig.~\ref{LuminosityFunctionFig2}. We quantitatively weighted the agreement between theory and observations by means of $\chi^2$ fits, from which it follows that the kinetic mixing parameter $\epsilon = (0.37^{+0.35}_{-0.37}) \times 10^{-14}$ at a 95\% confidence level ($i.e. \approx 2\sigma$-like), and the best fit value is $\epsilon = 0.372 \times 10^{-14}$ which lies in the 1.64$\sigma$ confidence region. We compared the anomalous cooling rate and WDLF data, and found that the WDLF 2$\sigma$ confidence as shown in Fig.~\ref{LuminosityFunctionFig3} ($i.e. \; \epsilon \leq 0.72 \times 10^{-14}$) is compatible with the cooling rate 2$\sigma$ confidence in Fig.~\ref{G117B15AFig} ($i.e. \; \epsilon = 0.97^{+0.35}_{-0.37} \times 10^{-14}$). Both approaches agreed with each other in confirming the existence of an anomalous rate of cooling of white dwarf with $\epsilon \simeq (0.60 \sim 0.72) \times 10^{-14}$ in this luminosity range. Our results also indicate that hidden photons are dominant in white dwarf radiations within the range of $6.8 \leq M_{Bol} \leq 10.5$, see Fig.~\ref{LuminosityFunctionFig1}. It is important to emphasize that both methods are complementary and equally sensitive to the emission of hidden photons in white dwarf.
\section*{ACKNOWLEDGMENTS}

We would like to thank Prof.$\;$Wah-Keung Sze and Prof.$\;$Georg G. Raffelt for useful communications. We also want to thank Robin Yu, Vincent Wu, Dr.$\;$Tandean and Dr.$\;$Mai for their suggestions and reminding. This work is supported by MOE Academic Excellent Program (Grant No:102R891505) and MOST of ROC.

\end{document}